\documentclass[twocolumn,secnumarabic,amssymb, nobibnotes, aps, prd]{revtex4-1}
\usepackage[british,UKenglish,USenglish,english,american]{babel}
\usepackage{graphicx}
\usepackage{dcolumn}
\usepackage{bm}
\usepackage[pdftex]{color}
\usepackage{tikz}        
\usetikzlibrary{decorations.pathmorphing}
\usetikzlibrary{patterns}
\usepackage{xcolor}
\usepackage{pgfplots}
\usepackage{float}
\usepackage[bookmarks,
colorlinks=true, linkcolor=black,
anchorcolor=black, citecolor=black, filecolor=black,
menucolor=black, pagecolor=black, urlcolor=black,
pdftitle={Full computational analysis of transient surface patterns in swelling hydrogels},
pdfauthor={Michele Curatolo and Paola Nardinocchi and Luciano Teresi and Eric Puntel}
]{hyperref}

\newcommand{\0} {\textbf{0}}

\newcommand{\dvg} {\texttt{div}\,}

\newcommand {\Bc}  {\mathcal{B}}

\newcommand {\Rc}  {\mathcal{R}}

\newcommand {\Tc}  {\mathcal{T}}

\newcommand {\eb} {\mathbf{e}}

\newcommand {\hb} {\mathbf{h}}

\newcommand {\mb} {\mathbf{m}}

\newcommand {\tb} {\mathbf{t}}
\newcommand {\ub} {\mathbf{u}}

\newcommand {\Bb} {\mathbf{B}}
\newcommand {\Cb} {\mathbf{C}}

\newcommand {\Fb} {\mathbf{F}}

\newcommand {\Ib} {\mathbf{I}}
\newcommand {\Mb} {\mathbf{M}}

\newcommand {\Pb} {\mathbf{P}}

\newcommand {\Sb} {\mathbf{S}}
\newcommand {\Tb} {\mathbf{T}}

\renewcommand{\b} {\textcolor{black}}
\definecolor{green}{RGB}{0,128,0}
\newcommand{\g} {\textcolor{black}}
\begin{document}
%
\title{Transient instabilities in swelling dynamics}%

\author{Michele Curatolo}%
\email[REVTeX Support: ]{michele.curatolo@uniroma3.it}
\affiliation{Universit\`a Roma TRE, Roma, Italy}
\author{Paola Nardinocchi}%
\email[REVTeX Support: ]{paola.nardinocchi@uniroma1.it}
\affiliation{Sapienza, Universit\`a di Roma, Roma, Italy}
\author{\g{Eric Puntel}}%
\email[REVTeX Support: ]{eric.puntel@uniud.it}
\affiliation{Universit\`a di Udine, Udine, Italy}
\author{\b{Luciano Teresi}}%
\email[REVTeX Support: ]{teresi@uniroma3.it}
\affiliation{Universit\`a Roma Tre, Roma, Italy}

\date{\today}

\begin{abstract}
We investigate the swelling dynamics driven by solvent absorption in a  hydrogel sphere immersed in a solvent bath, through an accurate computational model and numerical study. We extensively describe the transient process from  dry to wet and discuss the onset of surface instabilities through a measure of the lack of smoothness of the outer surface and a morphological pattern of that surface with respect to the two material parameters driving the swelling dynamics.
\end{abstract}

\maketitle
\tableofcontents

\section{Introduction}

\b{Hydrogels are soft materials made of cross--linked networks of hydrophilic polymers;
when immersed in water, they swell by absorbing the liquid until a new steady balance between elastic and chemical energy has been reached. The swelling--induced deformations may be very large, and that makes the mechanics of hydrogels especially interesting, and it also forces to set the stress--diffusion problem within the context of nonlinear mechanics 
\cite{Hong:2008,Duda:2010,Chester:2012,JMPS:2013,JAP:2015,SM:2015,Drozdov:2016,JAP:2016}.}

\b{The mechanics of hydrogels has attracted a lot of attention since decades \cite{Tanaka:1987,Hirotsu:1987,Suzuki:1997,Urayama:2012}. Stress diffusion modeling gives important explicit formulas describing
both the fast response of the hydrogels before diffusion starts, or the asymptotic response after 
diffusion-driven relaxation \cite{Urayama:2012,JAP:2016}. Nevertheless, explicit results describing the
transient dynamics are sought to find, while the numerical solutions of the stress diffusion model
is challenging; as already noted in \cite{Bertrand:2016}, transient dynamics received comparatively little attention despite its practical importance \cite{JAP:2014,SM:2014,Bouklas:2015,Drozdov:2016}.}
%

Whereas the surface instabilities which may characterise the steady state of hydrogels
being constrained in space and undergoing large volume variations have been largely studied \cite{Kang:2010,Weiss:2013,Shao:2016}, 
the same is not true in the case of transient surface instabilities. 
\b{An especially interesting transient phenomenon observed during the free swelling of hydrogels is the protrusion of surface patterns on the surface. 
It happens that, at early times, only a thin surface layer is swollen and the geometric mismatch
between this layer and the layers underneath may produce a sufficiently large pressures that 
make the outer surface to buckle.}
Surface patterns due to instability have been experimentally observed in both flat and non flat bodies \cite{Tanaka:1987,Barros:2012,Pandey:2013,Engelsberg:2013,Bertrand:2016}. On the other side, the theoretical and/or numerical characterization of the process is still lacking, even if a number of accurate studies have been proposed \cite{Bouklas:2015,Bertrand:2016}.

\b{Here, we study the swelling dynamics driven by solvent absorption in a  hydrogel sphere immersed in a solvent bath, by numerical experiments based on an accurate computational model;  in particular,
we observe and extensively describe the onset of surface instabilities.
The numerical experiments give insight into relevant quantities which are difficult or impossible to measure experimentally, as the stress state or the solvent concentration.} We also discuss the role of the two material parameters which completely drive the deformative process through a morphological phase diagram  which shows, for some choices of the two parameters, the morphology of the outer surface of the sphere.

\section{Theoretical background}
\label{sec:TB}
Our starting point is the multiphysics model presented and discussed in \citep{JMPS:2013} and successively refined in \citep{SM:2014}, where the buckling dynamics of a solvent--stimulated \g{and} stretched elastomeric sheet \g{are} investigated. 
\subsection{Displacement and solvent concentration}
We introduce a \emph{dry-reference state} $\Bc_d$  of the gel, and denote with $X_d\in\Bc_d$ a material point and with $t\in\Tc$ an instant of the time interval $\Tc$.
Our multiphysics model of gel has two state variables: the displacement field $\ub_d(X_d,t)$ ($[\ub_d]=$m), which determines the actual position $x$, at time $t$, of a point $X_d$ as $x=X_d+\ub_d(X_d,t)$,
 and the molar solvent concentration per unit dry volume $c_d(X_d,t)$ ($[c_d]=$mol/m$^3$).
\b{Key of the model is the volumetric constraint coupling the two state variables:}
\begin{equation}\label{Vconstraint}
J_d = \det\Fb_d = \hat J_d(c_d)=1+\Omega c_d\,,
\end{equation}
where $\Fb_d=\Ib+\nabla\ub_d$ is the deformation gradient and $\Omega$ is 
 the molar volume, that is, the volume per solvent mole ($[\Omega] =$ m$^3$/mol). The constraint (\ref{Vconstraint}) implies that any change in volume of the gel is accompanied by uptake or release of solvent.
\begin{figure}[h]
\centering
\includegraphics[scale=.65]{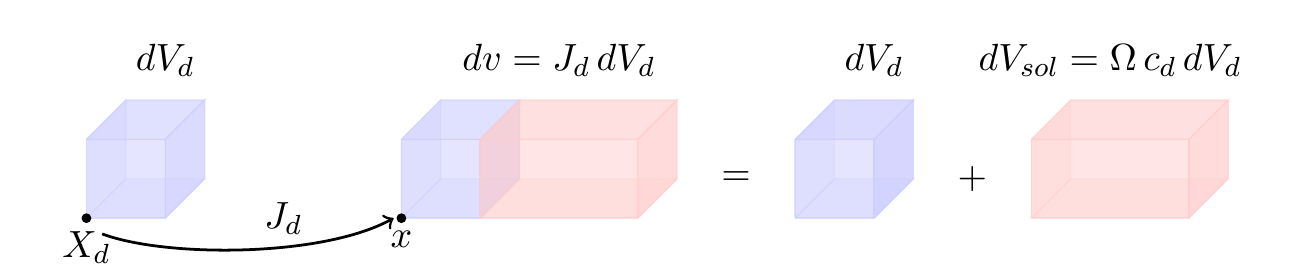}
\caption{\b{The volume change of volume elements, described by the Jacobian $J_d$, may be interpreted as
adding the elementary volume of solvent $\Omega\,c_d\,dV_d$ to the dry volume-element $dV_d$.}
\label{fig:0} }
\end{figure}
This in turn entails that the actual volume-element $dv$ of the body is
related to its dry volume-element $dV_d$ through the solvent concentration $c_d$, by the formula
\begin{equation}\label{Jconstraint}
\frac{dv}{dV_d}=J_d
                             = \hat J_d(c_d)=1+ \Omega c_d\,.
\end{equation}
The constitutive equation for the stress $\Sb_d$ ($[\Sb_d]$=Pa = J/m$^3$) at the dry configuration $\Bc_d$, henceforth termed dry--reference stress,
and for the chemical potential $\mu$ ($[\mu]$=J/mol) are derived from a relaxed version of the Flory--Rehner thermodynamic model \citep{FR1,FR2}. It is based on a free energy $\psi$  per unit dry  volume which depends on  $\Fb_d$  through an elastic component $\psi_e$, and on  $c_d$  through a polymer--solvent mixing energy $\psi_m$: $\psi=\psi_e+\psi_m$. The relaxed free--energy $\psi_r$ includes the volumetric constraint:
\begin{equation}\label{isoFR}
\psi_r(\Fb_d,c_d,p) = \psi_e(\Fb_d) + \psi_m(c_d) -p(J_d-\hat J(c_d))\,.
\end{equation}
The pressure $p$ represents the reaction to the volumetric constraint,
which maintains the volume change $J_d$ due to the displacement equal to the one due to solvent absorption or release $\hat J(c_d)$.
Key features of $\psi$ (or $\psi_r$) are the following: (\textbf{i}) $\psi$ is a density per unit volume of the dry polymer; (\textbf{ii}) the elastic contribution $\psi_e$ hampers swelling; (\textbf{iii}) the mixing contribution $\psi_m$ favors swelling.
\subsection{Stress and chemical potential}
The constitutive equations for the stress $\Sb_d$
and the chemical potential $\mu$ ($[\mu]$=J/mol) come from \b{dissipation issues} 
and prescribe that
\begin{equation}\label{csteqn}
\Sb_d=\hat\Sb_d(\Fb_d)-p\,\Fb_d^\star\,
\quad\textrm{and}\,\quad
\mu=\hat\mu(c_d)+p\,\Omega\,,
\end{equation}
with
\begin{equation}\label{mu_d}
\hat\Sb_d(\Fb_d)=\frac{\partial\psi_e}{\partial\Fb_d}\quad\textrm{and}\quad\hat\mu(c_d) = \frac{\partial\psi_m}{\partial c_d}\,,
\end{equation}
\g{where} $\Fb^\star = (\textrm{det}\,\Fb)\Fb^{-T}$. Typically, the Flory--Rehner thermodynamic model prescribes a neo-Hookean  elastic energy $\psi_e$ and a polymer--solvent mixing energy $\psi_m$:
\begin{equation}\label{nH}
\psi_e(\Fb_d) = \frac{G}{2}(\Fb_d\cdot\Fb_d-3)\,,\quad
\psi_m(c_d)= \frac{\g{\Rc} T}{\Omega}\,h(c_d)\,,
\end{equation}
with
\begin{equation}\label{FRm2}
h(c_d)=\Omega\,c_d\,\textrm{log}\frac{\Omega\,c_d}{1+\Omega\,c_d} + \chi\,\frac{\Omega\,c_d}{1+\Omega\,c_d}\,,
\quad [h]=1\,,
\end{equation}
\g{$G$ being} the shear modulus of the dry polymer, 
\b{$\g{\Rc}$ the universal gas constant, $T$ the temperature, and $\chi$ the Flory parameter.}
\b{Their physical units are $[G]$=J/m$^3$, $[\g{\Rc}]= $J/(K\,mol), $[T]=$ K, while 
$\chi$, called dis-affinity, is non dimensional and possibly temperature-dependent; its value is
specific of each solvent-polymer pair: high $\chi$ favours de--swelling, low $\chi$ drives swelling.}
\b{It is important to note that $G$ and $\Rc\,T/\Omega$ share the same physical dimensions:
they measure the volumetric density of the elastic and the mixing energy, respectively. 
The ratio $\varepsilon_m$ between elastic and chemical energy has an important role in swelling dynamics:} 
\begin{equation}
\b{\varepsilon_m=\frac{G\,\Omega}{\Rc\,T}\,.}
\end{equation}
\b{From (\ref{mu_d}) and (\ref{nH}) we obtain the constitutive equations for the dry-reference stress
$\hat\Sb_d(\Fb_d)$ and for chemical potential $\hat\mu(c_d)$;
this latter can be rewritten in terms of $J_d$ by exploiting the volumetric constraint (\ref{Vconstraint}):} %
\begin{equation}\label{mud}
\begin{array}{l}
\hat\Sb_d(\Fb_d)=G\,\Fb_d\,,\\[3mm]
\displaystyle{\hat\mu(c_d)=\hat\mu(J_d)=\Rc\,T\Bigl(\textrm{log}\frac{J_d-1}{J_d} + \frac{1}{J_d} + \frac{\chi}{J_d^2}\Bigr)\,.}
\end{array}
\end{equation}
\b{The actual stress (Cauchy) $\Tb$ is then given by the constitutive term $\hat\Tb(\Fb_d)$
minus the pressure term}
\begin{equation} \label{cauchy}
\b{\Tb=J_d^{-1}\Sb_d\,\Fb_d^T = \hat\Tb(\Fb_d) -p\,\Ib\,,}
\end{equation}
\b{with $\hat\Tb(\Fb_d)=G/J_d\,\Bb$, and $\Bb=\Fb_d\,\Fb_d^T$.}
%
\subsection{Solvent flux}
\label{sec:TB}
\b{A key element in the transient swelling is the solvent flux; 
here, we assume the following prescription for the reference solvent flux $\hb_d$}
\begin{equation}
\hb_d = \hb_d(\Fb_d,c_d,p) = -\Mb(\Fb_d,c_d)\nabla(\hat\mu(c_d)+p\,\Omega)\,.
\end{equation}
which is  consistent with the dissipation principle, provided that 
the mobility tensor \b{$\Mb(\Fb_d,c_d)$ is positive definite;
$[\Mb]$=mol$^2$/(s m J). Among the many admissible representations for the mobility,
here we assume $\Mb$ to be isotropic, and diffusion always to remain isotropic during any process (see Ref. \citep{JMPS:2013} for a full discussion on the different isotropic representations for $\Mb$), and linearly dependent on $c_d$:
We have:}
\begin{equation}
\Mb(\Fb_d,c_d) =  \frac{D}{\g{\Rc}T}\,c_d\Cb_d^{-1}\,,\quad
\Cb_d=\Fb_d^T\Fb_d\,,
\end{equation}
with $D$ ($[D]$=m$^2$/s) the diffusivity. Using $\mb$ to denote the outward unit normal,
$q=-\hb_d\cdot\mb > 0$ is a positive boundary source, that is, an inward flux.
\subsection{The Initial-Boundary Value problem}
The model is based on a system of bulk equations, describing the balance of forces
 and the balance of solvent concentration, coupled through the volumetric constraint (\ref{Vconstraint}), 
 and the constitutive equations (\ref{csteqn}): on $\Bc_d\times\Tc$
\begin{equation}\label{bal1}
0=\dvg\Sb_d \quad\textrm{and}\quad
\dot c_d = -\dvg\hb_d \,,
\end{equation}
with a dot denoting \g{the} time derivative and $\texttt{div}$ the divergence operator. 
Equations (\ref{bal1}) must be complemented with \b{mechanical boundary conditions on 
the traction $\tb$ and/or displacement $\bar\ub_d$:}
\begin{equation}\label{bal2}
\begin{array}{lll}
   \Sb_d\,\mb=\tb \,,   & \textrm{on } \partial_t\Bc_d\times\Tc\,,\\[3mm]
    \ub_d=\bar\ub_d\,,      & \textrm{on } \partial_u\Bc_d\times\Tc\,;
\end{array}
\end{equation}
and \b{with chemical boundary conditions on solvent source \b{$q_s$} and/or concentration $c_s$:}
\begin{equation}\label{bal3}
\begin{array}{ll}
   -\hb_d\cdot\mb = \b{q_s}\,,     & \textrm{on } \partial_q\Bc_d\times\Tc\,,\\[3mm]
   c_d =  c_s\,, & \textrm{on } \partial_c\Bc_d\times\Tc\,.
\end{array}
\end{equation}
Notation $\partial_s\Bc_d$ with $s = t,u,q$ or $c$ in the above equations denotes 
the portion of the boundary of $\Bc_d$ where traction $\tb$, displacement $\bar\ub_d$, flux $q_s$, and concentration $c_s$ are prescribed, respectively.
\b{Finally, the model is completed by the initial conditions 
for the state variables $\ub_d$ and $c_d$:}
\begin{equation}\label{bal4}
\ub_d = \ub_{do} \,,\,\, c_d=c_{do} \,,  \quad \textrm{on } \Bc_d\times\{0\}
\end{equation}
%
\section{Swelling equilibrium of a gel sphere}
\begin{figure}[h]
\centering
\includegraphics[scale=.8]{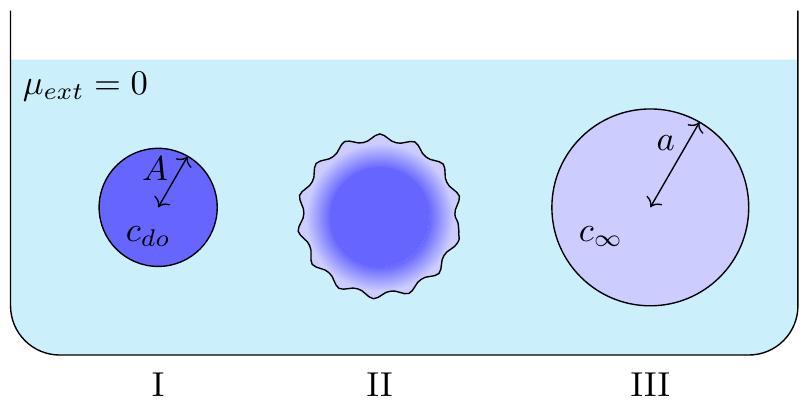}
\caption{\label{fig:1} From left to right: (I) the gel sphere of radius $A$ at the initial time; 
(II) at the early times; (III) at the final steady state of radius $a$.
\b{The blue coloring of the spheres points to different solvent concentration, uniform at initial and final states, and with a pronounced boundary layer at early times.} }
\end{figure}
\b{We now consider a spherical gel and reformulate the initial-boundary problem assuming radial symmetry,
that is, assuming $\ub_d=u_R(R,t)\,\mb$, with $R$ the radial coordinate and $\mb$ 
the unit radial vector, and $c_d=c_d(R,t)$.}
\b{Under these assumptions, the balance equations (\ref{bal1}) rewrites as}
\begin{equation}\label{bals}
S_R'+\frac{2}{R}\,(S_R-S_\theta)=0\,,\quad
\b{\dot{c_d}=-(h_R'+\frac{2\,h_R}{R})\,,}
\end{equation}
\b{with $S_R$ and $S_\theta$ the radial and hoop components of the dry--reference stress $\Sb_d$, $h_R$ the unique component of the flux vector $\hb_d(R,t)=h_R(R,t)\,\mb$, and a prime denoting  \g{derivation} with respect to the radial coordinate. On the boundary, we have zero radial stress $S_R=0$, and a 
concentration $c_s$ determined by the external chemical potential $\mu_{\rm ext}$, that is, 
equation $\hat\mu(c_s) + p\,\Omega=\mu_{\rm ext}$ holds.}
The constitutive equations determine the stress components
\begin{equation}\label{stress}
S_R= G\lambda_R-\lambda_\theta^2\,p\quad\textrm{and}\quad
S_\theta= G\lambda_\theta-\lambda_\theta\lambda_R\,p\,,
\end{equation}
as well as the flux field
\begin{equation}
h_R(R,t)=-\frac{D}{\Rc\,T}\,c_d(R,t)\lambda_R^{-2}(R,t)\mu'(R,t)\,,
\end{equation}
where $\lambda_R=r'$ and $\lambda_\theta=r/R$  are the radial and hoop stretches 
\b{and $r(R,t) = R + u_R(R,t)$ the actual radius.}
With this, equations (\ref{bals}) can be written as
\begin{eqnarray}\label{diffR}
\b{\dot c_d} &=& \frac{1}{R^2}\Bigl(\frac{D}{\Rc\,T}\;\frac{c_d}{\lambda_R^2}\;R^2
    \Bigl(\Rc\,T\;\frac{\partial h}{\partial c_d}c_d' + \Omega\,p'\Bigr)\Bigr)'\,,\nonumber\\[3mm]
0&=& G\lambda_R'-\lambda_\theta^2 p' \g{+}\frac{2}{R}G(\lambda_R-\lambda_\theta)\,.
\end{eqnarray}
Together with the incompressibility condition, which can be integrated and inverted to get
\begin{equation}\label{rdici}
r(R,t) = \Bigl(R^3 +3\;\Omega\int_0^R \varrho^2\;\b{c_d(\varrho,t)}\,d\varrho\Bigr)^{1/3}\,,
\end{equation}
\b{equations (\ref{diffR}) can be numerically integrated to describe swelling evolution under spherically symmetric conditions. }
\b{The steady fully-swollen state at $\mu_{\rm ext}=0$ is characterized by \g{an} uniform  concentration $c_\infty$ and an uniform swelling ratio $\lambda_R=\lambda_\theta=\lambda_\infty$,
with $\lambda_\infty=(1+\Omega\, c_\infty)^{1/3}$.}
\b{Such a state can be determined by solving the evolutive problem (\ref{diffR}).} 

\b{Alternatively,
both $\lambda_\infty$ and $c_\infty$ can be determined directly as solution of the steady problem: $\Sb_d=\0$ and $\mu=0$.
From (\ref{csteqn}), (\ref{mu_d}), and assuming that $\Fb=\lambda_\infty\,\Ib$, 
we have}
\begin{equation}\label{ss}
\textrm{log}(1-\frac{1}{\lambda_\infty^3}) +\frac{1}{\lambda_\infty^3} + \frac{\chi}{\lambda_\infty^6} + \frac{\varepsilon_m}{\lambda_\infty}=0\,.
\end{equation}
\b{Typically we have $\lambda_\infty>>1$} (that is, $1/\lambda_\infty<<1$), and equation (\ref{ss}) can be approximated  as
\begin{equation}\label{sper}
\varepsilon_m\lambda_\infty^8 + (\chi-1/2)\lambda_\infty^3 -\frac{1}{3}=0\,,
\end{equation}
\b{thus assuming the form of a singular perturbation. The leading order,
as already shown in \cite{Doi:2009,Dai:2011,PRSA:2014},  yields}
\begin{equation}\label{assln}
\lambda_\infty = \left(\frac{1-2\chi}{2\varepsilon_m}\right)^{1/5}\,.
\end{equation}
\b{It is worth noting that a scaling analysis based on:
i) the length scale $A$, 
with $A$ the radius of the sphere at dry-reference; 
ii) the characteristic time scale $t_c=A^2/D$; 
iii) the shear modulus $G$, shows that 
both the swelling dynamics and the steady solution are scale--free, and only depend on the  material parameters $\chi$ and $\varepsilon_m$.}

\b{However, as experiments showed \cite{Bertrand:2016}, swelling dynamics is not spherical symmetric 
at early and intermediate times, when wrinkles appear on the surface (see figure \ref{fig:1}).} 
Hence, we propose a refined computational analysis  based on the theoretical model shown in Section \ref{sec:TB} which allows us to highlight the pattern characteristics as well as the dependence on the material parameters.
%
\section{Finite Elements Analysis of Swelling Dynamics}
%
%

The Finite Elements Model (FEM) solves balance equations (\ref{bal1}) and the volumetric constraint (\ref{Vconstraint})
in a weak form as 
%
\begin{equation}\label{bala_weak1}
\begin{array}{l}
\displaystyle{
0=\int_{\Bc_d} -(\Sb_d(\Fb_d)-p\,\Fb_d^{\g{\star}})\cdot \nabla\tilde\ub_d\,,}\\[5mm]                    %
\displaystyle{
0= \int_{\Bc_d} [\, - \dot c_d \cdot\tilde c_d + \hb_d\cdot\nabla\tilde c_d]}\,, \\[5mm]
\displaystyle{
0=\int_{\Bc_d} [\, J_d - (1 + c_d\,\Omega\,)] \cdot \tilde p}\,,
\end{array}
\end{equation}
where the tilde indicates a test field.
We note that the unknown pressure $p$ is considered as an additional state variable, having the role of a Lagrange multiplier.
%
%

Boundary conditions (\ref{bal2}) are quite easy to handle, as we set $\tb=0$,
and assign a displacement $\bar\ub_d$ that eliminates any rigid motion without generating reaction forces.

Tackling the chemical boundary conditions (\ref{bal3}) is more tricky, as it is not possible to control the surface flux source $q_s$, nor the surface concentration $c_s$. Actually, what is done in real experiments, and what we aim at replicating in our numerical model, is the control of the chemical potential $\mu_{ext}$ of the bath on $\partial_c\Bc_d\times\Tc$.

In the present model, it is equation  (\ref{csteqn})$_2$, evaluated at the boundary, that relates
$c_s$ to $\mu_{ext}$; this is a highly non-linear equation which cannot be solved for $c_s$; moreover,
as we control the state variable $c_s$, the surface flux source $q_s$ must be considered as a 
reaction, which is
unknown a priori, and whose evaluation a posteriori yields poor approximations.

Those two issues are solved by posing in weak form both relation (\ref{csteqn})$_2$,
and the constraint (\ref{bal3})$_2$. 
%
\begin{equation}\label{bala_weak3}
0=\int_{\partial_c\Bc_d} [\,\mu(c_s)+p\,\Omega - \mu_{ext}\,]\cdot \tilde c_s\,,
\end{equation}
\begin{equation}\label{bala_weak4}
0=\int_{\partial_c\Bc_d} [\, (c_d-c_s)\,\g{\tilde q_s} + \g{q_s}\,(\tilde c_d-\tilde c_s)\,] \,\g{.}
\end{equation}
It is important to note that we use the same technique as before, that is, we enforce the constraint
$c_d=c_s$ by considering $q_s$ as an additional state variable, having the role of a Lagrange multiplier; as well known,
weak constraints provide a far better numerical evaluation of the boundary source $q_s$.

The complete problem can be reformulated as follows: find $\ub_d$, $c_d$, $p$,  $c_s$, and $q_s$ 
such that, for any test functions $\tilde\ub_d$, $\tilde c_d$, $\tilde p$, $\tilde c_s$, 
and $\tilde q_s$, equations (\ref{bala_weak1})--(\ref{bala_weak4}) hold; 
the three fields $\ub_d$, $c_d$, $p$ are defined in $\Bc_d\times\Tc$, while the two fields $c_s$ and 
$q_s$ are defined on $\partial_c\Bc_d\times\Tc$.
%
\subsection{General analysis of dynamics}
\b{We consider a sphere that is initially at equilibrium in nearly dry conditions, with
$\lambda_o\simeq 1.02$, corresponding to $c_{do}=1006$ mol/m$^3$ and $\mu_{ext}\simeq-6500$ J/mol} \footnote{If we assume that the sphere is in air of relative humidity $RH$ and that $\mu_{ext}=\g{\Rc}T\texttt{log}RH$, it means $RH\simeq 7\%$.}. 
\begin{figure}[H]
\centering
\includegraphics[scale=.7]{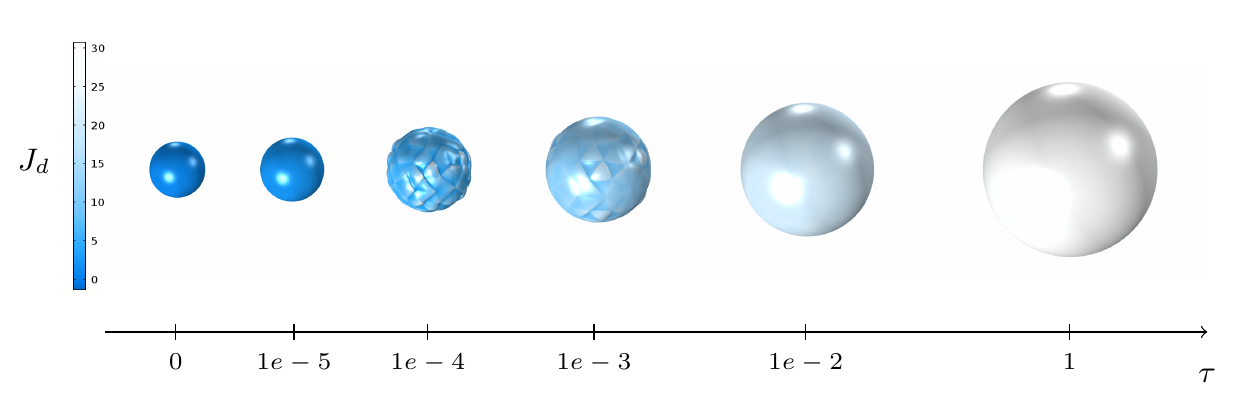}
\caption{\label{fig:2} Evolution of the surface patterns from the initial state, $\tau=0$, to the final steady state, $\tau=1$. Colormap shows the values of $J_d$.}
\end{figure}
\b{When the sphere is immersed in water at $\mu_{ext}=0$, it swells until a new spherical steady state is reached, having radius $a>A$. The steady fully-swollen state is characterized by \g{a} uniform  concentration field $c_\infty$ and a swelling ratio $\lambda_\infty$. With our choice of parameters, $\lambda_\infty\simeq 3.8$ (see Table \ref{tab:1}).}
\b{Let $\tau = t /(\alpha  t_c)$ be a non dimensional time;
we assign a time evolution law for the external chemical potential such that $\mu_{\rm ext}$ smoothly change 
from the initial value $\mu_{\rm ext}^o=-6500$ J/mol, to the final one $\mu_{\rm ext}^\infty=0$ J/mol, 
in a time interval $\tau_c<<t_c$.}
In particular, we define:
\begin{equation}
\mu_{\rm ext}(\tau) = \mu_{\rm ext}^o+(\mu_{\rm ext}^\infty-\mu_{\rm ext}^o)
(1-\texttt{exp}(-\tau/\tau_c))\,.
\end{equation}
\begin{figure}[H]
\centering
\includegraphics[scale=.7]{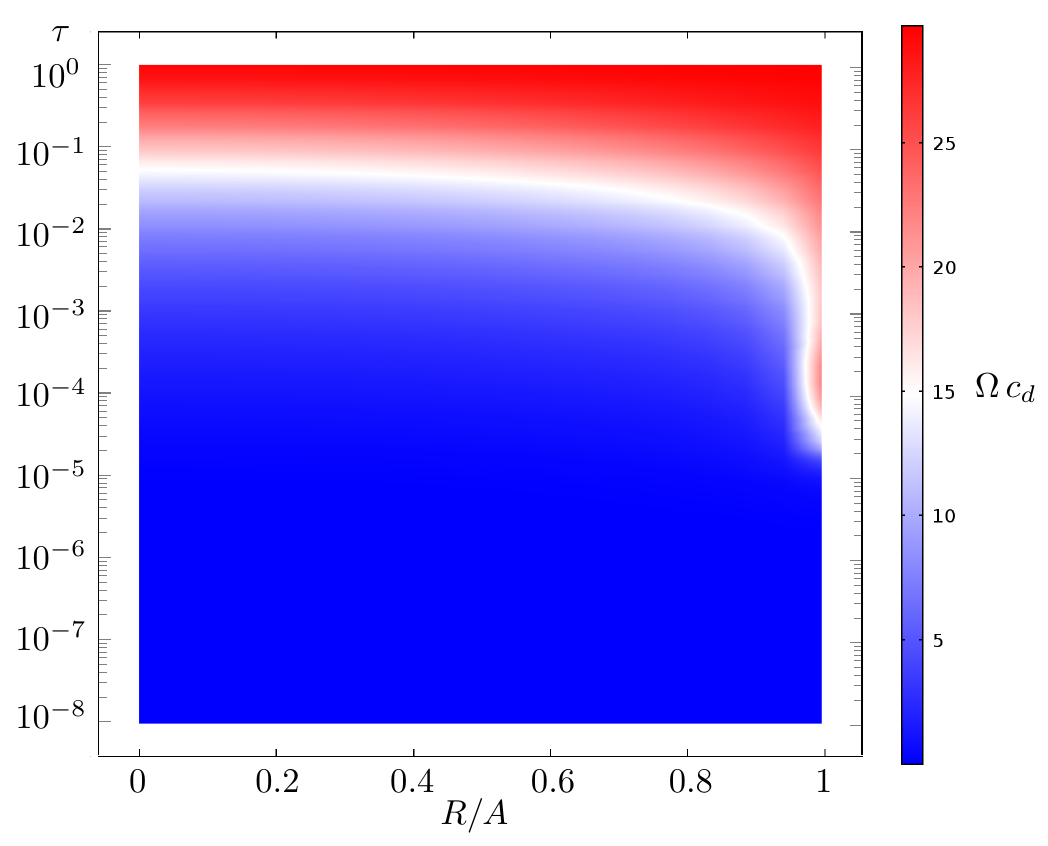}
\caption{\label{fig:31} 
\b{Contour plot of the volume change $\Omega\,c_d$ versus the 
dimensionless radius $R/A$, and time $\tau$ (in log scale); at early times, in the range 
$\tau\in( 1e^{-5}, 1e^{-2})$ it appears a thin layer at the outer surface, that is, solvent remains
confined in a small volume. Only for $\tau>1e^{-2}$ solvent enters deep in the gel, and becomes uniformly
distributed for $\tau\simeq 1$.
}}
\end{figure}
\begin{figure}[H]
\centering
\includegraphics[scale=.7]{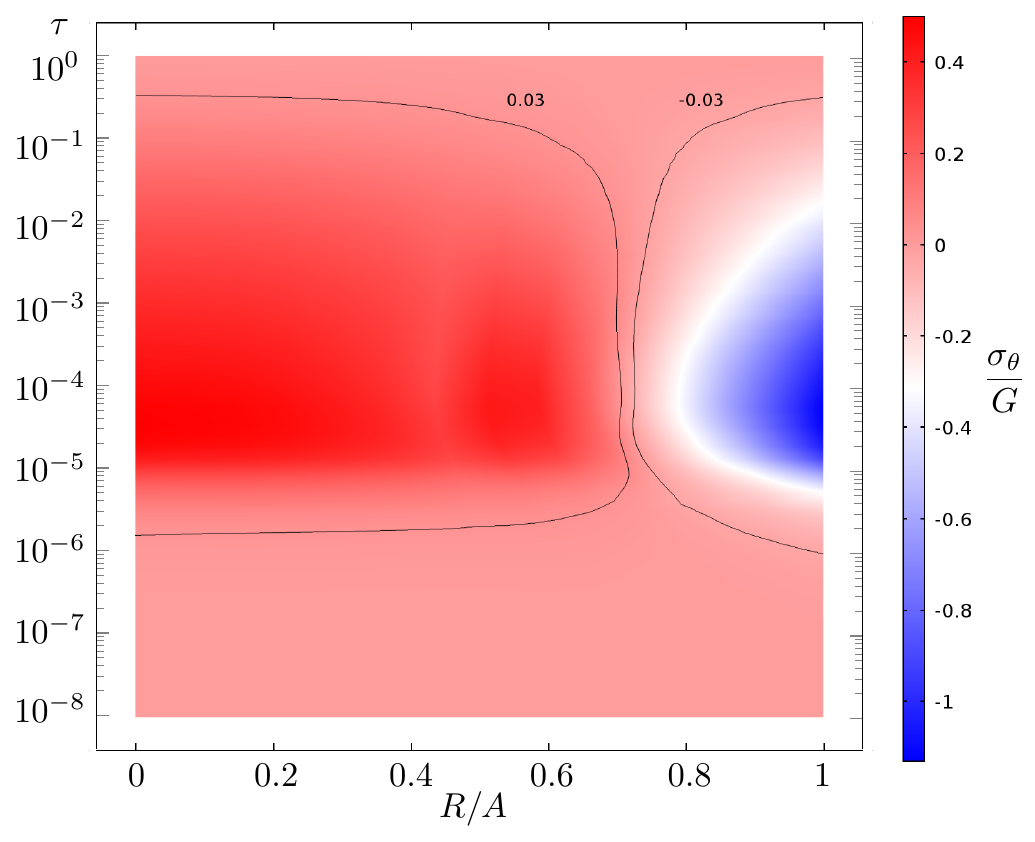}
\caption{\label{fig:32} 
\b{
Contour plot of the dimensionless hoop stress $\sigma_\theta/G$ versus the 
dimensionless radius $R/A$, and time $\tau$ (in log scale);
in the same time range $\tau\in( 1e^{-5}, 1e^{-2})$, a thin region under compression appears near the outer surface. Moreover, pressure rapidly changes from negative to positive when moving
towards the center of the sphere. Both features evidenced in the figures. \ref{fig:31} and \ref{fig:32} are the signature of wrinkling.}}
\end{figure}
%
\begin{table}[H]
\caption{Numerical values of the parameters}
\begin{ruledtabular}\label{tab:1}
\begin{tabular}{ll}
Parameter & Symbol and value\\
\hline
Shear modulus	& G = 50 kPa \\
Dis-affinity	& $\chi$  = 0.4 \\
Molar volume	& $\Omega$ = $1.8\times 10^{-5}$ m$^3$/mol\\
Time scaling    & $\alpha  = 3\times 10^4$ \\
\end{tabular}
\end{ruledtabular}
\end{table} 
\vskip-0.5cm\noindent
At $\tau=0$ the sphere is in almost dry conditions; at early times, $\tau<<1$, 
the swelling dynamics produces surface patterns which alters the spherical symmetry; 
such patterns disappear when the swelling evolves, and
before approaching the steady state, the gel completely recovers its smooth spherical shape (see figure \ref{fig:2}).
\b{Surface patterns are due to surface instabilities, which have been largely studied in growing soft materials \cite{BenAmar:2005,Goriely:2005,Dervaux:2011,Jia:2013}, even if, at the best of our knowledge, a 3D computational analysis based on a fully nonlinear multi-physics model of the swelling is still lacking.} Our model allows  to highlight the characteristics of swelling dynamics when surface instabilities appear, evolve, and disappear.
\b{At first, we define the hoop component $\sigma_\theta$ of the actual stress (Cauchy),
made of the constitutive part $\hat\sigma_\theta$, minus the indeterminate pressure $p$ 
\begin{equation}\label{hoopT}
\sigma_\theta=\hat\sigma_\theta - p\,,
\quad\textrm{with}\quad
\hat\sigma_\theta=\hat\Tb(\Fb_d)\,\eb_\theta\cdot\eb_\theta\,,
\end{equation}
where $\eb_\theta$ is a the unit vector orthogonal to $\mb$.} 

\b{According to the physical expectations and experimental observations \cite{Bertrand:2016}, 
at early times we observe a rapid swelling confined in a thin volume near the outer boundary, 
which is adjacent to an almost un-swollen core. }

\b{Figure \ref{fig:31} shows the contour plot 
of $\Omega\,c_d$ versus $R/A$ and $\tau$: at early times, in the range $\tau\in( 1e^{-5}, 1e^{-2})$
it appears a thin boundary layer (red colored) where $\Omega\,c_d$ is quite higher with respect 
the values it attains in the un-swollen core (blue colored)}. 
\b{Analogously, figure \ref{fig:32} shows that, within the same time interval, 
the boundary layer is under negative hoop stress (compressive stress), meanwhile in the larger zone beneath
the boundary layer, the hoop stress is positive (tensile stress).}

\b{With the contour plot of $\sigma_\theta/G$ we can track the time evolution of the boundary between the compressive and the tensile regions, and verify the relationship between zones having high solvent gradient 
and compressive hoop stress.}
\begin{figure}[H]
\centering
\includegraphics[scale=.8]{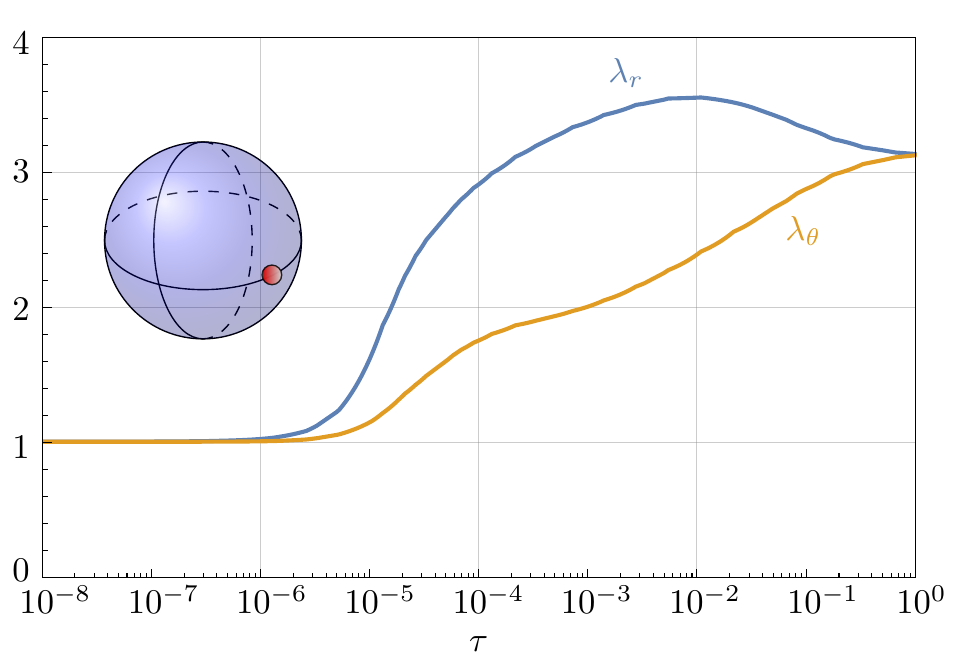}
\caption{\label{fig:7} \b{Time evolution of the hoop $\lambda_\theta$ and radial $\lambda_R$ stretches
at a point on the outer surface of the sphere (semi-logarithmic representation).}}
\end{figure}
\begin{figure}[H]
\centering
\includegraphics[scale=.75]{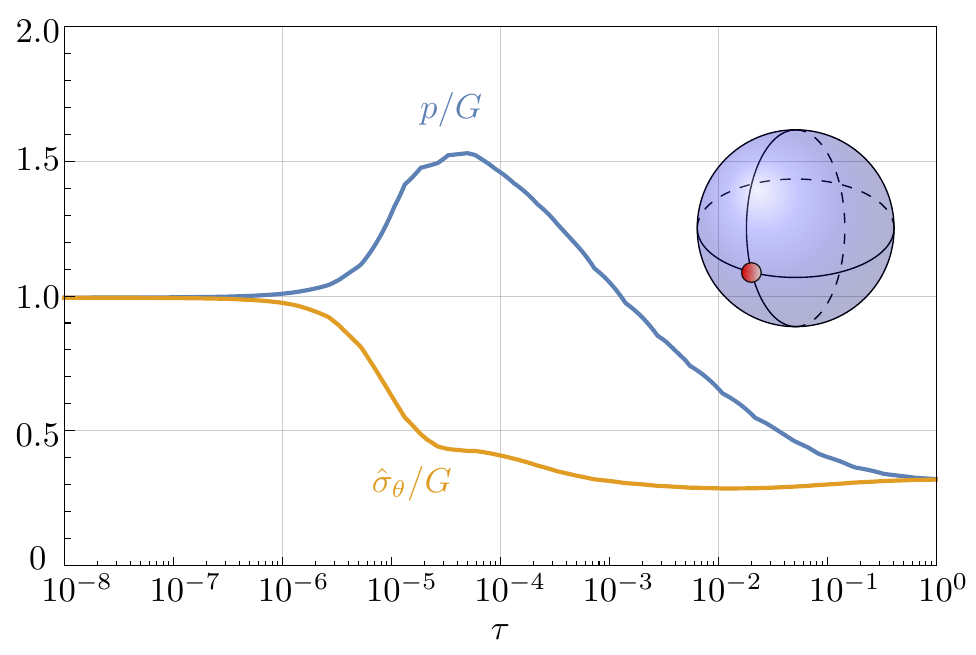}
\caption{\label{fig:56} 
\b{Time evolution of the dimensionless pressure $p/G$ (blue) and of the dimensionless
hoop stress $\hat \sigma_\theta/G$ (orange) at a point on the outer surface; compare
with plots in Fig. \ref{fig:56c}; semi-logarithmic representation.}}
\end{figure}
\begin{figure}[H]
\centering
\includegraphics[scale=.75]{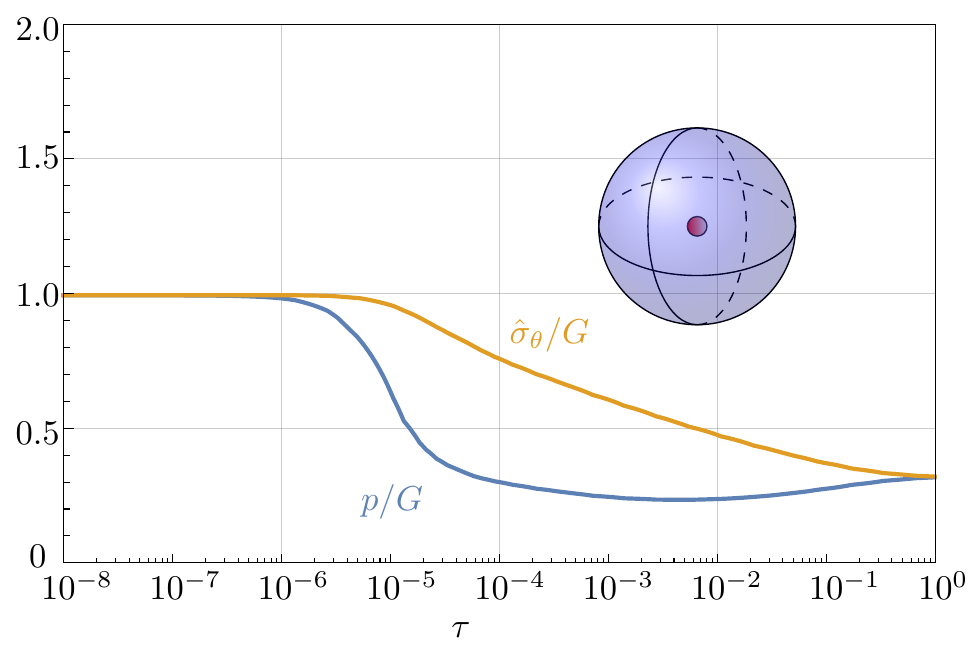}
\caption{\label{fig:56c} 
\b{Time evolution of the dimensionless pressure $p/G$ (blue) and of the dimensionless
hoop stress $\hat \sigma_\theta/G$ (orange) at a point on the outer surface
; compare with plots in Fig. \ref{fig:56}; semi-logarithmic representation.}}
\end{figure}
\b{It is useful to represent separately the two terms that add up to make the dimensionless hoop stress 
$\sigma_\theta/G$, see (\ref{hoopT}): the first one $\hat \sigma_\theta/G$ is constitutively determined,
and known as the effective stress in poro-mechanics;
the second one $p/G$ represents the mechanical contribution to the dimensionless chemical potential as equation (\ref{csteqn})$_2$ shows} 

\b{Figures \ref{fig:56} and \ref{fig:56c} show that $p/G$ has a key role in determining the compressed region: 
while the effective stress $\hat \sigma_\theta/G$ is monotone decreasing both at the boundary and inside
the sphere, the pressure behaves very differently.
On the boundary of the sphere, $p/G$ increases very fast from the initial value, and remains always greater than $\hat \sigma_\theta/G$ (top panel); conversely, at the center 
$p/G$ decreases very fast from the initial values, and remains always smaller than $\hat \sigma_\theta/G$
(bottom panel). The two summands of the stress attain the same value at $\tau=1$, which correspond to a steady, stress-free state.}

So, the stress state of the sphere is similar to the one we find in a circumferentially growing thick shell due to the residual stresses triggered by growth: tensile in the inner layer and compressive in the outer one \cite{BenAmar:2005}. Likewise,  both hoop and radial strains $\lambda_\theta$ and $\lambda_R$ are always larger than $1$ on the outer surface,  and drive the swelling-induced hoop and radial growth of the sphere (see figure \ref{fig:7}). 
\b{We can evaluate the volume $V_s$ of solvent crossing the boundary during the dimensionless time 
interval $(0,1)$, as well as its time rate $\dot V_s$, that is, the volume of solvent crossing the boundary per unit of dimensionless time:}
\begin{equation}
V_s=\int_0^1 \dot V_s \,\g{t_c\,d\tau}\,\quad\textrm{and}\quad
{\dot V_s}=\Omega\,\int_{\partial\Bc_d} \g{q_s} \,dA_d\,,
\end{equation}
\b{being $\Omega\,q_s$ ($[\Omega\,q_s]$ = m$^3$/(s$\cdot$m$^2$))} the volume of solvent crossing the boundary per unit time and unit area. Figure \ref{fig:8} shows that the rate $\dot V_s$ is especially high at early times.
\begin{figure}[h]
\centering
\includegraphics[scale=.8]{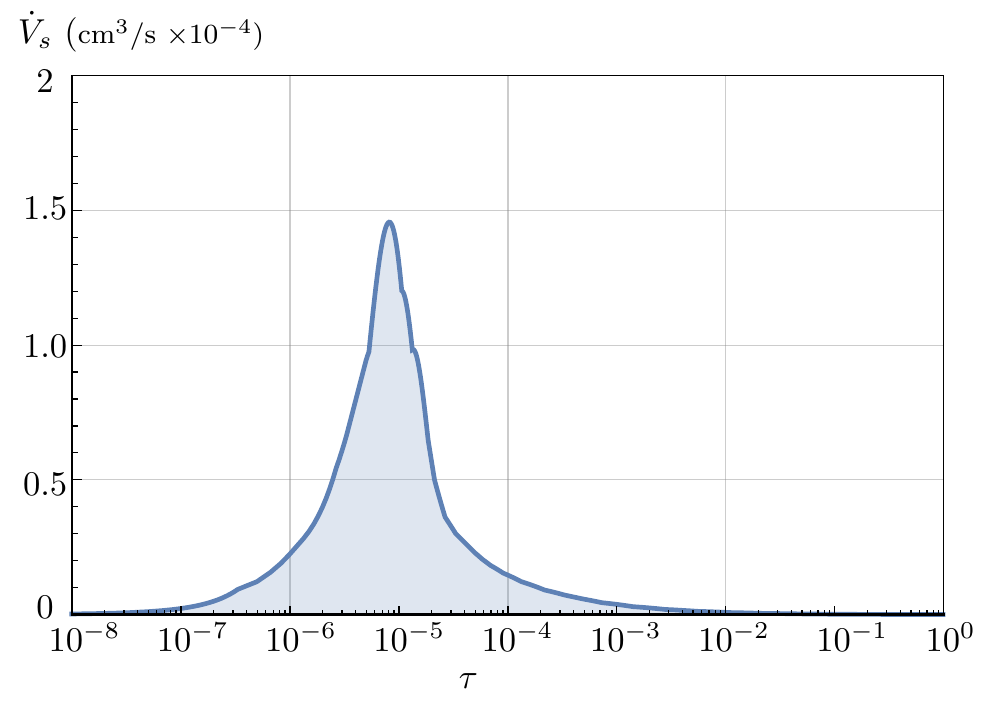}
\caption{\label{fig:8} 
\b{The volume rate of solvent uptake $\dot{V}_s$ reaches a peak value at early times (semi-logarithmic representation).}}
\end{figure}
\begin{figure}[h]
\centering
\includegraphics[scale=.8]{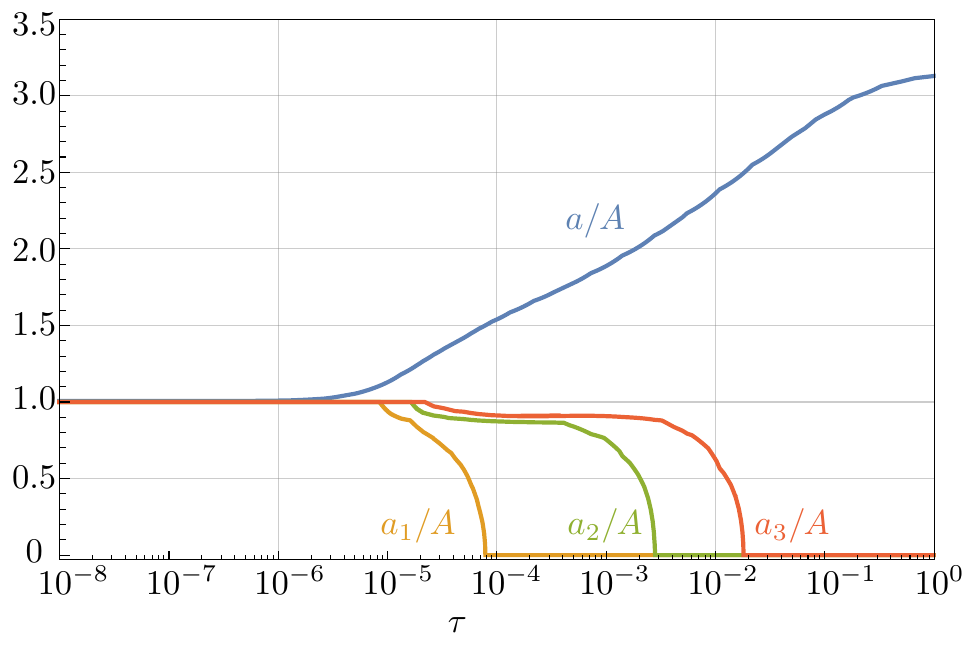}
\caption{\label{fig:8} 
\b{Time evolution of the dimensionless radius $a/A$ compared to the evolution of the three
dimensionless radii $a_i/A$ of the contour levels where solvent uptake has value
$\Omega\, c_{d1}=1$, $\Omega\,c_{d2}=5$, and $\Omega\, c_{d3}=10$ (semi-logarithmic representation).}
}
\end{figure}
\b{Finally, we show the evolution of the dimensionless radius $a_i/R$ ($i=1,2,3$),
with $a_i$ the radius of the contour level of the solvent uptake $\Omega\,c_{di}$; 
with set $\Omega\, c_{d1}=1$, $\Omega\,c_{d2}=5$, and $\Omega\, c_{d3}=10$.} Interestingly, as we expected and in contrast with the measurements made via a shadowgraph technique and presented in \cite{Bertrand:2016}(Appendix 1), the dimensionless radius \b{$a_i/A$ always decreases for any choices of the threshold value $\Omega\,c_{di}$.}
\subsection{Surface instabilities}
To quantify the bumpiness of the spherical surface 
we introduce two different measures, one based on the surface area, the other on the surface gradient of the displacement.
\begin{figure}[]
\centering
\includegraphics[width=8.5cm]{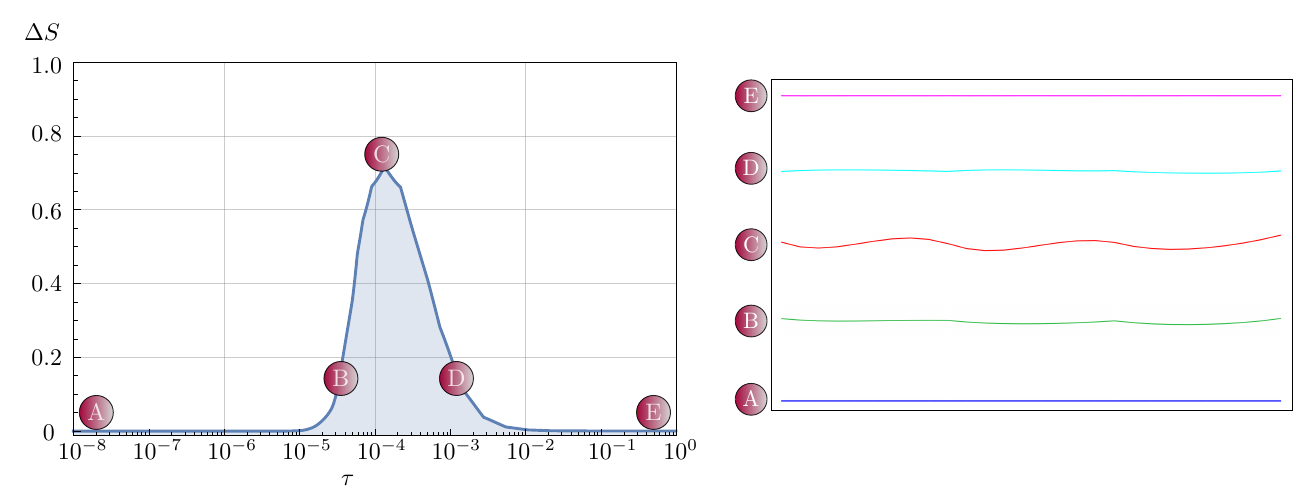}
\caption{\label{fig:10}
Left: Time evolution of $\Delta{S}=S_{wr} - S_{sm}$ (semi-logarithmic representation).
Right: Wrinkled pattern of the surface corresponding to the points $A,B,C,D,E$ of the left panel.}
\end{figure}
\b{The actual, non dimensional area $S_{wr}$ of the spherical gel, be it wrinkled or not, is given by} 
\begin{equation}
S_{wr} = \frac{1}{4\pi A^2}\,\int_{\partial\Bc_d}\vert\Fb_d^\star\mb\vert\, dA_d \,,
\quad \Fb_d^\star = J_d\Fb_d^{-T}\,,
\end{equation}
\b{being  $\vert\Fb_d^\star\mb\vert$ the ratio between the swollen area element and the corresponding 
area element $dA_d$ of the dry surface. Then, we introduce the non dimensional area $S_{sm}$ of the mean sphere}
\begin{equation}
S_{sm} = \frac{4\pi\bar{r}^2}{4\pi A^2}\,,\quad
\bar{r}=\frac{1}{4\pi A^2}\,\int_{\partial\Bc_d}r(t)\,dA_d \,,
\end{equation}
\b{being $\bar r$ the average radius of the actual outer surface}. 
The evolution of the difference $\Delta{S}=S_{wr} - S_{sm}$ shows a peak during the critical time interval, when surface instabilities attain their maximum value, as figure \ref{fig:10} shows (left panel), together with a qualitative view of surface profile (right panel).
\begin{figure}[]
\centering
\includegraphics[scale=.8]{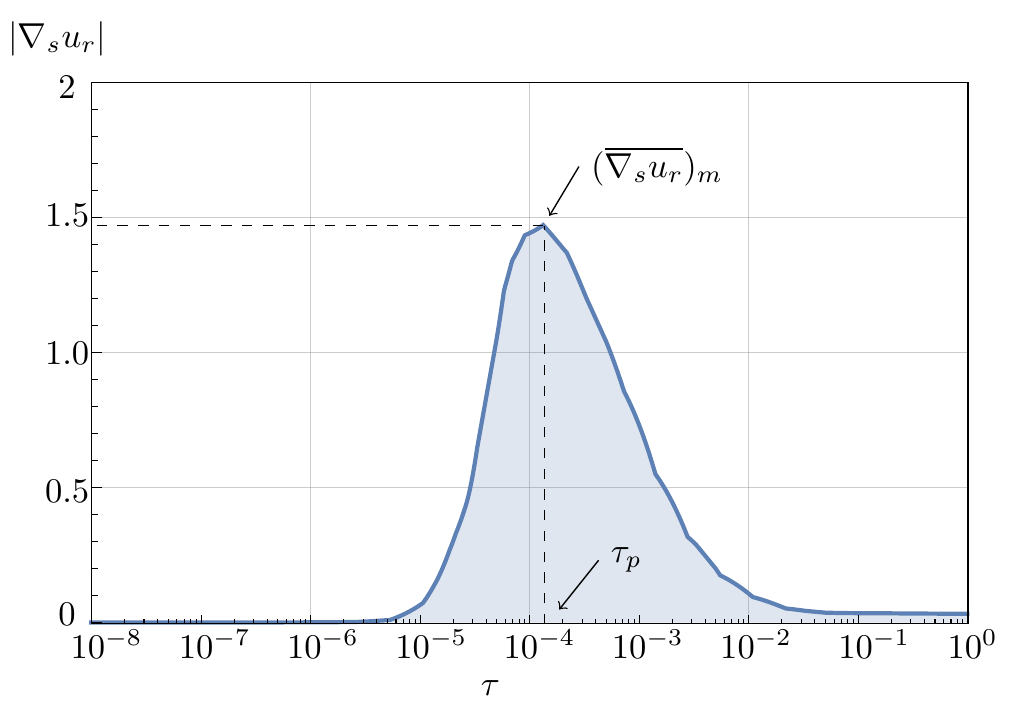}
\caption{\label{fig:11} Evolution of the module $\vert \nabla^s u_r\vert$ of the surface gradient taking the maximum value $\vert \nabla^s u_r\vert_m$ (semi-logarithmic representation).}
\end{figure}
\b{The other measure of the bumpiness of the external surface is based on the surface gradient
$\nabla_s u_R$ of the radial displacement $u_R$. Given the surface projector 
$\Pb_s=(\Ib-\mb\otimes\mb)$, we define}
\begin{equation}
\nabla_s u_r=\Pb_s\nabla u_r\,,\quad
\overline{\nabla_s u_r}=\frac{1}{4\pi A^2}\int_{\partial\Bc_d}\|\nabla_s u_r\|\,dA_d\,,
\end{equation}
\b{where $\|\nabla_s u_r\|=(\nabla_s u_r \cdot \nabla_s u_r)^{1/2}$ is the norm of $\nabla_s u_r$.
The mean surface gradient $\overline{\nabla_s u_r}$ is not monotone in time and has its maximum 
$(\overline{\nabla_s u_r})_{\rm max}$ at $\tau_p$, as figure \ref{fig:11} shows.}
\begin{figure*}[h]
\centering
\includegraphics[scale=.9]{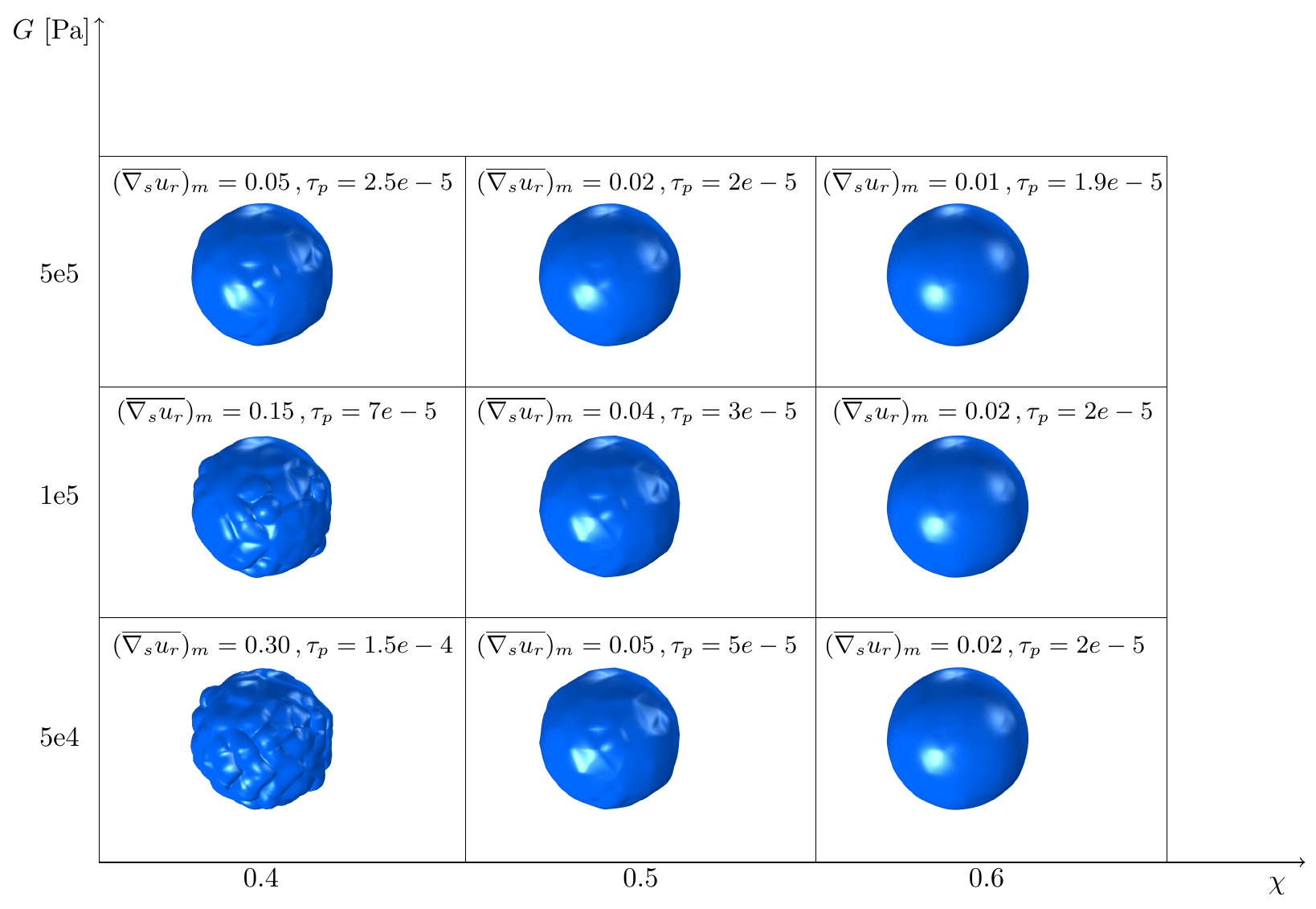}
\caption{\label{fig:12} Morphological phase diagrams of emerging surface patterns from numerical simulations. The different surface morphologies are displayed for varying values of the material parameters $G$ and $\chi$. The value $\vert \nabla^s u_r\vert_m$ as well as the time $\tau_p$ when that value is attained are shown for each pair of material parameters.}
\end{figure*}

\b{To assess the influence of the two material parameters $G$ and $\chi$ on the bumpiness of the sphere,
we ran a series of analyses with $G=(0.5e5{\rm Pa}, 1e5{\rm Pa}, 5e5{\rm Pa})$, 
and $\chi=(0.4, 0.5,0.6)$. }
We present the result of our analyses through a morphological phase diagram showing the patterns on the outer surface of the sphere, \b{the value of $(\overline{\nabla_s u_r})_{\rm max}$ and of $\tau_p$ for each choice of the two parameters. As expected, the minimum value of $(\overline{\nabla_s u_r})_{\rm max}$} corresponding to an almost smooth outer surface is attained for the highest values of $G$ and $\chi$, both determining a reduced swelling due to the high elastic stiffness $G$ of the polymeric network and to a larger dis--affinity $\chi$ between solvent and polymer.

\section{Conclusions}
We presented an extensive computational study of the swelling dynamics driven by solvent absorption in a  hydrogel sphere immersed in a solvent bath, based  on a fully three--dimensional nonlinear stress--diffusion model. In particular,
we observed and  described the onset of surface instabilities, introducing appropriate measures of the surface bumpiness.
To catch surface patterns due to the high and fast swelling in the the thin surface outer layer, we formulated the boundary conditions on solvent flux and concentration in a  form of weak constraints, so providing a  better numerical evaluation of the boundary flux which is the determinant of the surface instabilities.

The analysis gives insight into relevant quantities which are difficult or impossible to measure experimentally, as the stress state or the solvent concentration, also showing key differences with other characteristics wrinkling patterns observed during swelling--induced growth.

\begin{acknowledgments}
M.C., E.P., and L.T. acknowledge the National Group of Mathematical Physics (GNFM--INdAM) for support.
\end{acknowledgments}

\end{document}